\documentclass[aps,prl,twocolumn,groupedaddress,showpacs]{revtex4}
\usepackage[french]{babel}
\usepackage{graphicx}
\usepackage{color}
\usepackage{amsmath}
\usepackage{amssymb}

\begin{document}
\title{Visibility of a Bose-condensed gas released from an optical lattice at finite temperatures}
\author{Fabrice Gerbier}
\affiliation{Laboratoire Kastler Brossel, D{\'e}partement de
Physique de l'ENS, 24, rue Lhomond, 75 005 Paris, France}
\author{Simon F{\"o}lling, Artur Widera, and Immanuel Bloch}
\affiliation{Institut f{\"u}r Physik, Johannes
Gutenberg-Universit{\"a}t, 55099 Mainz, Germany.}
\date{\today}
\begin{abstract}
In response to a recent manuscript \cite{diener2006a} on the
analysis of interference patterns produced by ultracold atoms
released from an optical lattice, we point out that in the presence
of a Bose-Einstein condensate the interference pattern can be
strongly modified by interaction effects and the presence of a
harmonic trap superimposed on the lattice potential. Our results
show that the visibility of the interference pattern is significant
only if a sizeable condensate fraction is present in the trap, in
strong contrast to the findings of ref.~\cite{diener2006a}.
\end{abstract}
 \maketitle
%
%
%
\paragraph{Introduction-}
Experiments with ultracold quantum gases in optical lattices rely
heavily on time-of-flight expansion to probe the spatial coherence
properties of the trapped gas
\cite{greiner2002a,stoeferle2004a,gerbier2005a,gerbier2005b}. When
the phase coherence length is large compared to the lattice spacing,
the post-expansion density distribution reveals a striking
interference pattern with the same symmetry as the reciprocal
lattice. As the phase coherence length decreases, {\it e.g.} on
approaching the Mott insulator transition, the visibility of this
interference pattern decreases accordingly, and for very short
coherence length the underlying lattice order can only be revealed
through higher-order correlations \cite{foelling2005a}. A recent
manuscript \cite{diener2006a} has raised the question, whether even
a minute condensate fraction on the order of $1/(\epsilon N^{2/3})$
($N$ being the total particle number) should give rise to an
interference pattern with visibility $\mathcal{V}$ very close to
unity, $\mathcal{V}\approx 1 - \epsilon$. According to the
calculations in \cite{diener2006a}, a system of, {\it e.g.},
$N=10^5$ atoms should show an interference pattern with a visibility
of $\mathcal{V} \approx > 99\%$ for minute condensate fractions
$\approx 4\%$ ($\approx 5000$ atoms). As current experiments with
ultracold quantum gases do not observe this (the best visibilities
that have been achieved in our group being close to 90\%), the
authors conclude that the current experiments must be carried out in
a thermal regime with no condensate or superfluid fraction present
in the trap.

In this paper we would like to point out that this reverse
conclusion is an artifact of the model used in \cite{diener2006a} to
describe the time-of-flight expansion. A central assumption in Ref.
\cite{diener2006a} is to consider that the condensate momentum
distribution is Heisenberg-limited by the size of the system prior
to expansion, and that this distribution is preserved in
time-of-flight. Below, we would like to explain that several effects
spoil this assumption, as under realistic experimental conditions,
one needs to take into account:

\begin{itemize}
    \item (i) The inhomogeneous trapping potential for the quantum gas in the
    lattice,
    \item (ii) The mean-field broadening of the momentum peaks during time-of-flight
    expansion,
    \item (iii) The finite time-of-flight in the experiments.
\end{itemize}

Practically, these effects tend to broaden the peaks present in the
interference pattern, hence diminishing its peak value. In the
presence of a significant non-condensed fraction, this will also
lower the visibility. In cases when the initial momentum width is
very large, for example for a cloud close to or in the Mott
insulator regime, or a thermal gas well above the critical
temperature, they can usually be neglected. However, the momentum
width for a condensate is conversely very narrow. Even without an
optical lattice, effects (i)-(iii) have been known to play an
important role in the interpretation of time-of-flight images (see,
{\it e.g.} \cite{kagan1996a,castin1997a,dalfovo1999a}). We point out
that a number of other effects might be relevant in a given
experimental situation, for instance a finite imaging resolution, or
collisions during the initial expansion phase that would scatter the
atoms out of the condensate. Here we do not consider these aspects
further, and consider an idealized experiment limited only by the
three points above. Below we outline our arguments and point out the
most critical steps to arrive at the conclusions given in
\cite{diener2006a}. We propose a simple model to explain why the
visibility of the interference pattern follows the condensate
fraction in the system at least in a qualitative way. We also find
that for the parameters that correspond to current experiments, the
visibility of an expanding, non-condensed thermal cloud is always
significantly lower than $1$, a conclusion consistent with
experiments
\cite{greiner2002a,stoeferle2004a,gerbier2005a,gerbier2005b}.

\paragraph{Setup and Optical Lattice Potentials -}
The optical lattice potential can be expressed as
\begin{equation}
V_{\rm OL}({\bf r})=V_0\left( \sin^2(k_{\rm L}x)+\sin^2(k_{\rm
L}y)+\sin^2(k_{\rm L}z)\right).
\end{equation}
Here $V_0$ is the lattice depth expressed in units of the
single-photon recoil energy $E_{\rm R}=h^2/2m\lambda_{\rm L}^2$.
Here $k_{\rm L}=2\pi/\lambda_{\rm L}$ is the laser wavevector,
$\lambda_{\rm L}$ is the laser wavelength and $m$ is the atomic
mass. In addition to the lattice potential, an ``external''
potential is present, due to the presence of a magnetic trap and
also to the optical confinement provided by the Gaussian-shaped
lattice beams. The external potential is nearly harmonic with
trapping frequency
\begin{equation}\label{omega}
\omega_{\rm T}\approx\sqrt{\omega_{\rm m}^2+\frac{4(2
V_0-\sqrt{V_0})}{m w^2}},
\end{equation}
where $\omega_{\rm m}$ is the frequency of the magnetic trap,
assumed isotropic, and where $w$ is the waist ($1/e^2$ radius) of
the lattice beams, assumed identical for all axes. This formula,
slightly different from the one given in \cite{gerbier2005b}, takes
into account the modification of the vibrational ground state energy
in each well due to the spatial variations of the laser intensities
on the scale $w$ (see \cite{greinerphd}), which amounts to a few
percents change in $\omega_{\rm T}$.

After releasing the cloud from the trap, the density distribution of
the expanding cloud after a time of flight $t$ is usually taken to
be proportional to the momentum distribution,
\begin{equation}\label{ntof}
n({\bf k})= \prod_{i=x,y,z} \left|\tilde{w}\left(k_i
\right)\right|^2 \mathcal{S}\left({\bf k}\right),
\end{equation}
with a scaling factor ${\bf r}_{\rm t.o.f.}=(\hbar t/m){\bf k}$.
Eq.~(\ref{ntof}) is valid provided that two conditions are met:
interactions have a negligible influence on the expansion, and the
time of flight is long enough to treat the initial particle
distribution as point-like (``far-field'' approximation). The
envelope $\left|\tilde{w}\right|^2$ is the Fourier transform of the
Wannier function in the lowest Bloch band, which we approximate by a
Gaussian, $\left|\tilde{w}(k_i)
\right|^2\approx\frac{w_0}{\pi^{1/2}}\exp{\left(-k_i^2
w_0^2\right)}$, with $w_0$ the size of the on-site Wannier function.
$S({\bf k})$ is given by $\mathcal{S}({\bf k})= \sum_{i,j} e^{i{\bf
k}\cdot({\bf r}_i-{\bf r}_j)}\langle \hat{a}_i^\dagger
\hat{a}_j\rangle$. Experimentally, one records the momentum
distribution integrated over the probe direction $z$, which we call
$n_\perp$. We assume for simplicity that the width of
$\mathcal{S}\left({\bf k}\right)$ is lower or comparable to $1/w_0$
(this should hold in the situations where the population in the
first excited Bloch band is negligible). Then, substituting
$\left|\tilde{w}\left( k_z \right)\right|^2$ with
$\left|\tilde{w}\left(0 \right)\right|^2$ leads to
\begin{equation}\label{ntof2}
n_\perp({\bf k}_\perp)\approx  A d \int dk_z \mathcal{S}\left({\bf
k}\right),
\end{equation}
where we introduced the lattice spacing $d$, a global factor
$A=\frac{w_0}{\sqrt{\pi}d} \left|\tilde{w}\left(k_x \right)\right|^2
\left|\tilde{w}\left(k_y \right)\right|^2$. The visibility as
defined in \cite{gerbier2005a} follows from
\begin{equation}\label{V}
\mathcal{V}=\frac{n_{\rm max}-n_{\rm min}}{n_{\rm max}+n_{\rm min}},
\end{equation}
where $n_{\rm max}= n_0({\bf 0})+n_{\rm th}({\bf 0})$ is measured at
the first lateral peaks of the interference pattern, {\it e.g.} at
${\bf K}_{\rm max}=(1/2,0)\times 2\pi/d$, and where the minimum
density $n_{\rm min}= n_{\rm th}({\bf K}_{\rm min})$ is measured
along a diagonal with the same distance from the central peak, {\it
i.e.} at the point defined the vector ${\bf K}_{\rm
min}=(1/\sqrt{2},1/\sqrt{2})\times 2\pi/d$.

\paragraph{Orders of magnitude -}
When condensed and non-condensed components coexist, the planar
momentum distribution $n_\perp$ splits naturally into a condensate
contribution $n_0({\bf r}_\perp)$ and a non-condensed contribution
$n_{\rm nc} ({\bf r}_\perp)$. Let us first consider the relative
scaling of the condensate and non-condensate contributions. We write
the condensate momentum distribution, characterized by the condensed
atom number $N_0$ and a momentum width $\Delta k_0$, as
\begin{equation}
n_0({\bf k}_\perp)= A \frac{N_0}{(\Delta k_0 \cdot d)^2} f_0 \left(
\frac{ | {\bf k} |}{\Delta k_0} \right),
\end{equation}
where $f_0$ is a periodic, dimensionless function. Similarly,
introducing the number of of non-condensed atoms $N_{\rm nc}$, we
write the non-condensed contribution as
\begin{equation}
n_{\rm nc}({\bf k}_\perp)= A \frac{N_{\rm nc}}{(\Delta k_{\rm nc}
\cdot d)^2} f_{\rm nc} \left( \frac{ | {\bf k} |}{\Delta k_{\rm nc}}
\right).
\end{equation}
We assume the condensate contribution to produce a sharp
interference pattern ($f_0({\bf K}_{\rm max})=1$ and $f({\bf K}_{\rm
min})\approx 0$), whereas the non-condensed part extends over the
whole Brillouin zone ($\Delta k_{\rm nc} \approx 1/d$). The
visibility then reads
\begin{equation}\label{V}
\mathcal{V}=\frac{N_0 + \left( \Delta k_0 d\right)^2 N_{\rm nc}~
f^{(-)} }{N_0 + \left( \Delta k_0 d\right)^2 N_{\rm nc}~ f^{(+)}}
\end{equation}
where we introduced
\begin{equation}
f^{(\pm)}=\left( f_{\rm nc}({\bf K}_{\rm
max})\pm f_{\rm nc}({\bf K}_{\rm min})\right)/f_0({\bf K}_{\rm max})
\end{equation}
which take values of order unity. The ``suppression factor'' $\left(
\Delta k_0 d \right)^2$ sets the importance of the non-condensed
fraction; if this ratio is very small, then even a tiny condensed
number will yield a visibility close to unity.

\begin{figure}
\includegraphics[width=8.7cm]{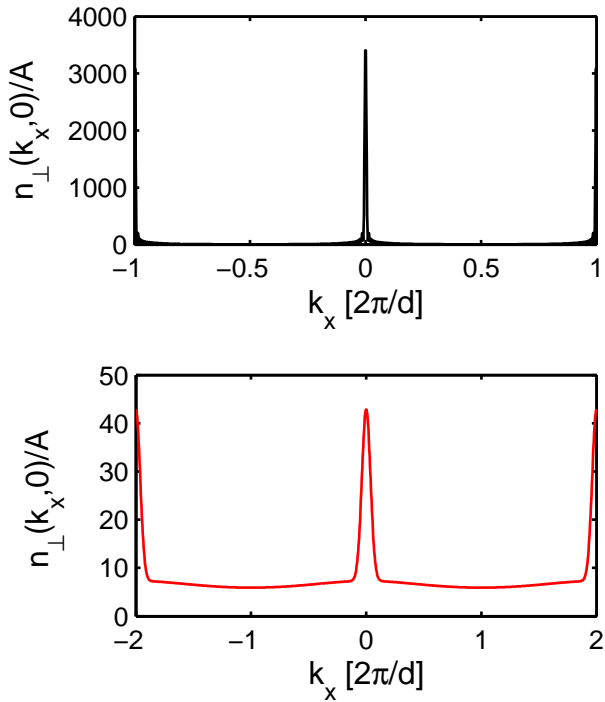}
\caption{Planar momentum distribution for $k_y=0$. The upper panel
is plotted for a uniform ideal gas with $N_s=10^6$ sites and one
atom per site, and for a condensed fraction $f_c\sim 5$~\%. The
lower panel is plotted for our harmonic trap model with interactions
(using $\alpha=0$), also for a condensed fraction $f_c\sim 5$~\% and
a total number $N=10^6$ using the trap parameters given in
\cite{gerbier2005b}. Note the vastly different scales on the
vertical axes.} \label{graphnk}
\end{figure}

\paragraph{Uniform system -}
The system considered in \cite{diener2006a} is a gas in a box
with $N_s$ lattice sites in each direction ($L$ in the notation of
Ref.~\cite{diener2006a}). The momentum width for the condensate is
diffraction limited to $\Delta k_0 \sim 1/N_s d$. Essentially this
amounts to the assumption that the {\it post-expansion} momentum
distribution of the condensate is a delta function. In
\cite{diener2006a}, any structure in the non-condensed component is
neglected, so that $f$ is uniform. Then,
\begin{equation}\label{Vuniform}
\left( \Delta k_0 d\right)^2 \sim \frac{1}{N_s^2} \sim 10^{-4}
\end{equation}
for a typical $N_s\sim 100$. This is where the result in Eq.~(4) of
Ref.~\cite{diener2006a} comes from, up to a factor $w_0/d$ which
depends on how the $k_z$ integral is handled. In Fig.~\ref{graphnk}
(upper panel) the momentum distribution for such a case is
displayed. Essentially, the momentum peaks approach delta functions,
explaining why tiny condensate fractions can lead to visibilities
very close to unity. We note that a better way of analyzing the
visibility, is to measure the weight, i.e. the integral over a peak
over a finite area, which accounts for finite resolution effects
that are inevitably present in experiments.

\paragraph{Ideal gas in a harmonic trap -}
As mentioned above, current experiments are carried out in the
presence of a harmonic trap superimposed on the lattice. In this
case, one expects the ground state wavefunction to be localized on a
scale given by the harmonic potential ground state extension
$\sigma_{\rm T}=\sqrt{\hbar/m\omega_{\rm T}}$, and not by the
``number of lattice sites'' (roughly equivalent to the size $w_0$ of
the laser beams). In fact, usually one has $w_0 \gg \sigma_{\rm T}$.
The main point here is that the suppression factor is no longer
hopelessly small, especially when most atoms are non-condensed.
However, as well known, this ideal gas model for the condensate is a
poor description of realistic experimental situations
\cite{dalfovo1999a}.

\paragraph{Interacting gas in a harmonic trap -}

To obtain a better description, it is necessary to include
interatomic interactions in the picture. In the Thomas-Fermi limit,
which usually applies to experimental situations, the in-trap
spatial condensate wavefunction will be significantly broader than
in the ideal gas case, implying a narrower momentum distribution.
However, as argued above, the post-expansion momentum profile can be
very different from the sharply peaked in-trap profile. First, the
interaction energy is released as kinetic expansion energy, and
second, for a finite time-of-flight $t$ the initial size of the
distribution is not necessarily negligible. Due to these effects,
the actual momentum distribution will differ from the one calculated
using Eq.~(\ref{ntof}).

Here, we handle them in a heuristic way, by approximating the maxima
in the planar momentum distribution for the condensate as Gaussians,
\begin{equation}
n_0\left({\bf k}_\perp \approx {\bf K}_{\rm max}\right) \approx
\frac{N_0 \pi^2 A}{(\Delta k_{\rm eff} d)^2}\exp\left(-{\bf
k}_\perp^2/\Delta k_{\rm eff}^2 \right).
\end{equation}
We assume an effective spatial width given by the quadratic sum of
the initial width $\sigma_0$ and of the expansion width $\hbar
\Delta k_0/m t$ corresponding to the release energy. In momentum
space, this corresponds to $\Delta k_{\rm eff}^2 =\Delta k^2
+\left(\frac{m \sigma_0}{\hbar t}\right)^2$. In the Thomas-Fermi
limit, both the post-expansion momentum width $\Delta k$ and the
initial size $\sigma_0$ are determined by the interaction energy,
which itself is proportional to the chemical potential $\mu$.
Accordingly, one has
\begin{eqnarray}
\frac{\hbar^2\Delta k^2}{m}& =\alpha \mu, ~~ m \omega_T^2 \sigma_0^2
&  =\beta \mu,
\end{eqnarray}
where the coefficients $\alpha$ and $\beta$ are of order unity. The
final result is
\begin{eqnarray}\label{dk_eff2}
\Delta k_{\rm eff}^2 \approx\frac{m
\mu}{\hbar^2}\left(\alpha+\frac{\beta}{(\omega_T t)^2} \right),
\end{eqnarray}
corresponding to a suppression factor $\left( \Delta k_{\rm eff}
d\right)^2$ proportional to $\frac{m d^2 \mu}{\hbar^2}\sim \mu/E_R
\lesssim 1$ much larger than expected from Eq.~(\ref{Vuniform}) (we
estimate below the proportionality factor). Hence, one expects the
observed visibility to follow the evolution of the condensed
fraction at least qualitatively. The argument also holds for the
quantum-depleted part of the non-condensed cloud.

\paragraph{Visibility for a trapped gas model}
To make the preceding arguments more precise, we introduce a
semiclassical description of the thermal cloud, for which
interaction effects can usually be neglected to a first
approximation \cite{dalfovo1999a}. The thermal component is
characterized by a distribution function in phase space given by
\begin{eqnarray}
\label{rhoth} \rho({\bf r},{\bf k}) &
=\frac{1}{\exp\left[(\epsilon_{{\bf k}}+V(\bf{r})-\mu)/k_{\rm B} T
\right]-1},
\end{eqnarray}
with the tight-binding dispersion relation $\epsilon_{{\bf k}}  = 2J
\sum_{i=x,y,z}\left(1-\cos(k_i d)\right)$, with the trap potential
$V({\bf r})=\frac{1}{2}m \omega_T^2 r^2$. We can obtain the thermal
fraction by integration over phase space,
\begin{eqnarray}\label{Nth}
N_{\rm th} & = (2\pi)^{3/2} \left(\frac{R_{\rm th}}{d}\right)^3
\sum_{n=1}^\infty \frac{e^{n (\mu-6 J)/k_{\rm B} T}}{n^{3/2}}
I_0\left(2 n J/k_{\rm B} T\right)^3.
\end{eqnarray}
Here $I_0$ is a Bessel function and $ R_{\rm th} = \sqrt{\frac{k_B
T}{m \omega_T^2}}$ the thermal cloud radius. The critical
temperature for condensation follows from Eq.~(\ref{Nth}) by setting
the chemical potential $\mu=0$ \cite{dalfovo1999a}. Integrating the
phase space distribution over position and $k_z$ gives the planar
momentum distribution,
\begin{eqnarray}\nonumber
n_\perp({\bf k}_\perp) & = & A (2\pi)^{3/2} \left(\frac{R_{\rm
th}}{d}\right)^3  \sum_{n=1}^\infty \frac{e^{n (\mu-2 J)/k_{\rm B}
T}}{n^{3/2}}
\\ \label{nperp_th} & & I_0\left(2 n  J/k_{\rm
B} T\right) e^{-n \epsilon_{{\bf k}_\perp}/k_{\rm B} T}.
\end{eqnarray}
This takes the form announced in Eq.~(\ref{V}).

\begin{figure}
\includegraphics[width=8.7cm]{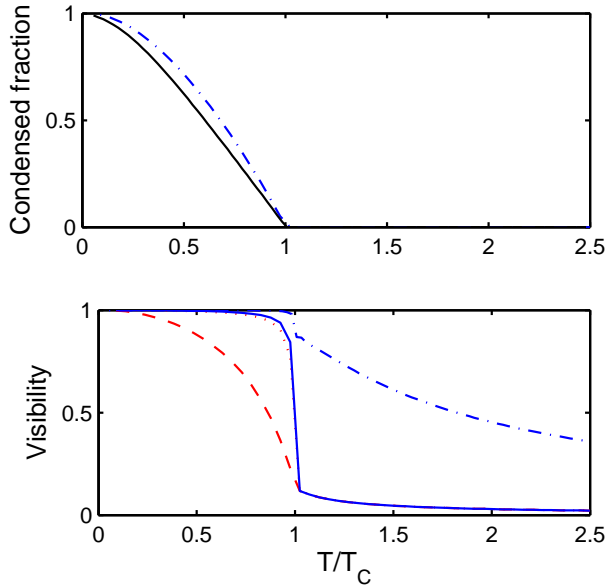}
\caption{Upper plot: Calculated condensed fraction for an ideal Bose
gas in a periodic potential, with and without additional harmonic
trap. Lower plot: visibility of the interference observed after
release from the trap. The dash-dotted line corresponds to an
uniform ideal gas, as in the model of \cite{diener2006a}, the solid
line to a harmonically trapped ideal gas. The long and short-dashed
lines correspond to two models of an interacting condensate, which
should be seen as upper and lower bounds for the visibility
($\alpha=0$ and $\alpha=8/21$, see text). The curves are plotted for
$N=10^6$ atoms, $V_0=10~E_R$ and an expansion time $t=20~$ms.}
\label{graphV}
\end{figure}

To estimate $\alpha$ and $\beta$ in Eq.~(\ref{dk_eff2}), we impose
that our Gaussian model should reproduce the same interaction and
potential energy as a condensate in the Thomas-Fermi limit, with
chemical potential given by \cite{greinerphd}
\begin{eqnarray}
\mu=\left(\frac{15}{16\sqrt{2}\pi} N_0 U\right)^{2/5}\left(m
\omega_T^2 d^2\right)^{2/5}.
\end{eqnarray}
The potential energy per particle \cite{dalfovo1999a} $E_{\rm
pot}=\frac{3}{7}\mu=\frac{3}{4}m\omega_T^2 \sigma_0^2$ gives a
coefficient $\beta \approx 12/21$. The coefficient $\alpha$ depends
on how the interaction energy is redistributed during the expansion.
We consider the two limits, where the interaction energy is entirely
concentrated either into the single diffraction peak under
consideration ($E_{\rm int}=2\mu/7=3\hbar^2 \Delta k_0^2/4m$, or
$\alpha=8/21$), or in the central peak avoided by the visibility
measurement ($\alpha=0$). This provides upper and lower bounds for
the actual visibility. The results of the calculations is plotted in
Fig. \ref{graphV} for various assumptions: no trap, ideal gas in a
trap, interacting gas in a trap with $\alpha=0$ or $\alpha=8/21$.
Comparing the two latter cases, one sees that the final result
depends sensitively on the value of $\alpha$. A more complete
calculation would probably yield a curve lying in between the two
bounds plotted. Nevertheless, irrespective of the particular model,
the curves demonstrate that the visibility falls well below unity in
a broad temperature range, where the condensate fraction stays
non-zero.

In conclusion, we have calculated the visibility of the interference
pattern observed when releasing a finite temperature, Bose-condensed
gas from an optical lattice. Using a realistic model for the
condensate momentum distribution, we find that the evolution of the
visibility with temperature or lattice depth follows the evolution
of the condensed fraction at least qualitatively. We therefore
disagree with the conclusion made in Ref.~\cite{diener2006a}, that a
measured visibility deviating from unity necessarily implies the
absence of a Bose condensate in the system. This claim originates
essentially from neglecting interaction effects on the gas
expansion, which is a valid approximation close or in the Mott
insulator regime, but a poor one for a system including a sizeable
condensate, with or without \cite{dalfovo1999a} an optical lattice.
We do however agree with the authors of \cite{diener2006a} on the
importance of finite temperature effects in current experiments, as
implied by recent experimental
\cite{stoeferle2004a,gerbier2006a,foelling2006a} and theoretical
work
\cite{demarco2005a,rey2006a,lu2006a,capogrosso2006a,gerbier2007a},
and certainly consider this topic to warrant further investigations.

We would like to thank Martin Zwierlein for useful comments on the
manuscript.



\end{document}